\documentclass[11pt,a4paper,useAMS,usenatbib]{emulateapj}
\usepackage{epsfig}
\usepackage{amsmath}
\usepackage{natbib}

\begin{document}

\title{Detection of a Luminous Hot X-ray Corona \\ Around the Massive Spiral Galaxy NGC266}

\author{\'Akos Bogd\'an\altaffilmark{1}, William R. Forman, Ralph P. Kraft, and Christine Jones}
\affil{Smithsonian Astrophysical Observatory, 60 Garden Street, Cambridge, MA 02138, USA; abogdan@cfa.harvard.edu}
\email{$^1$Einstein Fellow}

\shorttitle{LUMINOUS X-RAY CORONA AROUND NGC266}
\shortauthors{BOGD\'AN ET AL.}

\begin{abstract}
The presence of luminous hot X-ray coronae in the dark matter halos of massive spiral galaxies is a basic prediction of galaxy formation models. However, observational evidence for such coronae is very scarce, with the first few examples having only been detected recently. In this paper, we study the large-scale diffuse X-ray emission associated with the massive spiral galaxy NGC266. Using \textit{ROSAT} and \textit{Chandra} X-ray observations we argue that the diffuse emission extends to at least $\sim$$70$ kpc, whereas the bulk of the stellar light is confined to within $\sim$$25$ kpc. Based on X-ray hardness ratios, we find that most of the diffuse emission is released at energies $\lesssim$$1.2$ keV, which indicates that this emission originates from hot X-ray gas. Adopting a realistic gas temperature and metallicity, we derive that in the $(0.05-0.15)r_{200}$ region (where $r_{200}$ is the virial radius) the bolometric X-ray luminosity of the hot gas is $(4.3\pm0.8)\times10^{40}\ \rm{erg \ s^{-1}}$ and the gas mass is $(9.1\pm0.9)\times10^{9}\ \rm{M_{\odot}}$. These values are comparable to those observed for the two other well-studied X-ray coronae in spiral galaxies, suggesting that the physical properties of such coronae are similar. This detection offers an excellent opportunity for comparison of observations with detailed galaxy formation simulations. 
\end{abstract}


\section{Introduction}
The existence of gaseous X-ray coronae around massive galaxies was first predicted by \citet{white78}. Following this work, galaxy formation simulations aimed to characterize such coronae \citep{white91,toft02,crain10,vogelsberger12}. The continous advances in the implemented physics and applied methodology in the simulations led to major revisions in the predicted properties of the coronae. Given the fundamental nature of these coronae and the disagreement between models, the observational characterization of luminous X-ray coronae offers a unique insight to probe  the physical processes that influence galaxy formation and evolution. 

Although hot gaseous coronae are ubiquitous for massive early-type galaxies \citep[e.g.][]{forman85,osullivan01,bogdan11}, these coronae cannot be used to probe structure formation models. On the one hand, the observed coronae around early-type galaxies may not originate from the accretion of primordial gas, but from the merger process and the corresponding starburst. On the other hand, massive early-type galaxies tend to lie in rich environments, and hence the coronal X-ray gas generally cannot be distinguished from the larger-scale emission associated with the group/cluster. Thus, galaxy formation models can  be best tested by exploring hot X-ray coronae around massive spiral galaxies.

The hunt for luminous X-ray coronae \textit{beyond} the optical radii of massive spiral galaxies began with \textit{ROSAT}    and continued with \textit{Chandra} and \textit{XMM-Newton} observations \citep[e.g.][]{benson00,rasmussen09}. However, early studies failed to detect hot coronae around normal (undisturbed and low star formation rate) spirals, and only upper limits could be derived. Recently, \citet{anderson11} detected a luminous X-ray corona around NGC1961, and  \citet{dai12} found a hot  corona around UGC12591. They measured the spectrum of the integrated emission within 40 kpc for NGC1961 and  24  kpc for UGC12591.  While the resulting spectra were dominated by the X-ray emission associated with the galaxies, the integrated temperatures ($0.60^{+0.10}_{-0.09}$ keV and $0.64\pm0.03$ keV) were consistent with the expectations for a hot corona. In \citet{bogdan13} we derived the coronal temperatures and metallicities of NGC1961 and NGC6753 beyond their optical radii, specifically in the $23.5-70.5$ kpc and $22.0-66.0$ kpc range. In these regions, the gas temperatures are $0.61^{+0.10}_{-0.13}$ keV for NGC1961 and  $0.69^{+0.06}_{-0.07}$ keV for NGC6753, which are independent of the X-ray emission associated with the stellar body of the galaxies.  

At present only three non-starburst spiral galaxies are detected with luminous X-ray coronae. Given the fundamental importance and the extreme rarity of these coronae, it is essential to identify other systems and explore the variations in coronal properties as a function of the host galaxy properties. Characterizing the coronae in a broad sample of galaxies will place stringent constraints on galaxy formation models. 

\begin{figure*}[t]
  \begin{center}
    \leavevmode
      \epsfxsize=5.8cm\epsfbox{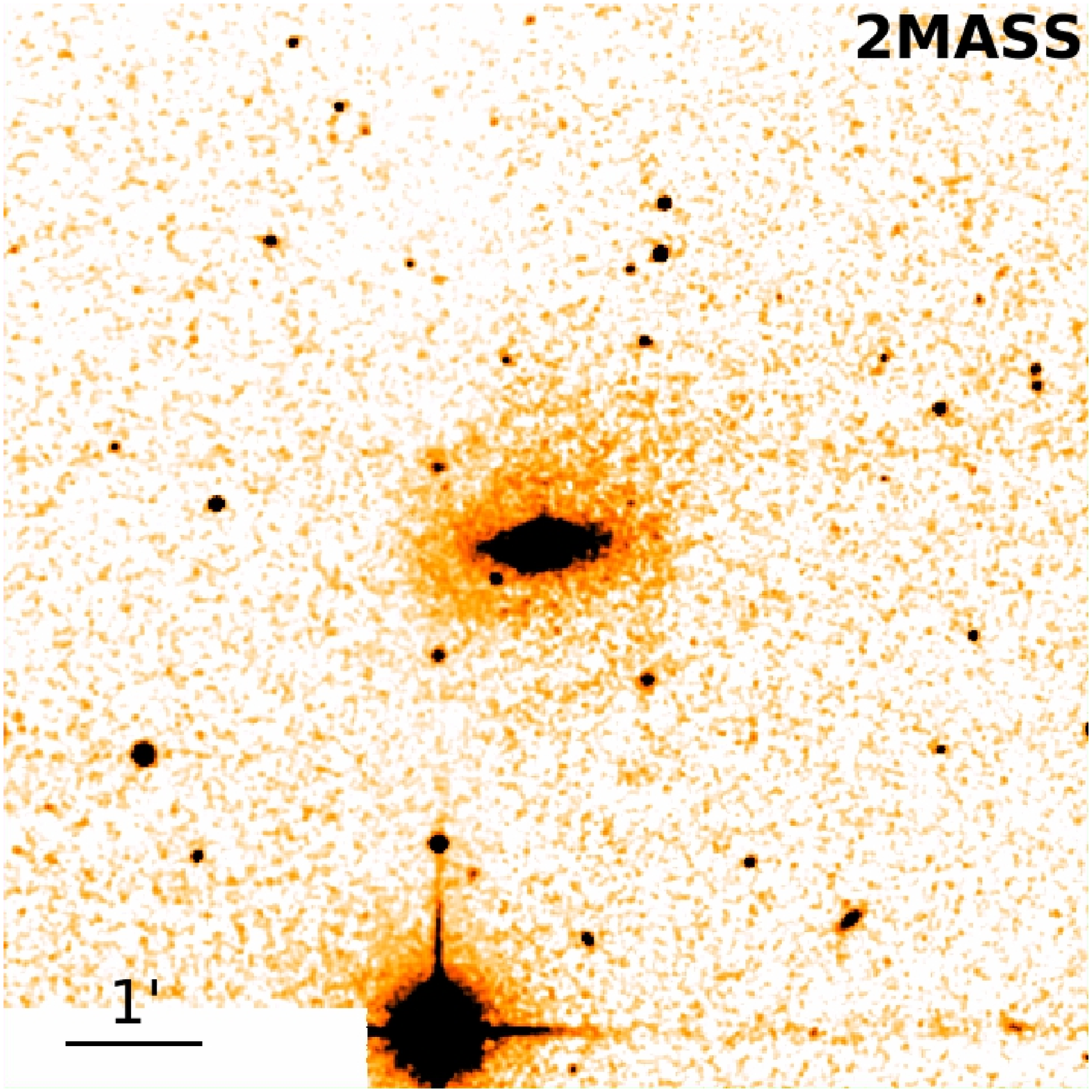}
\hspace{0.1cm} 
      \epsfxsize=5.8cm\epsfbox{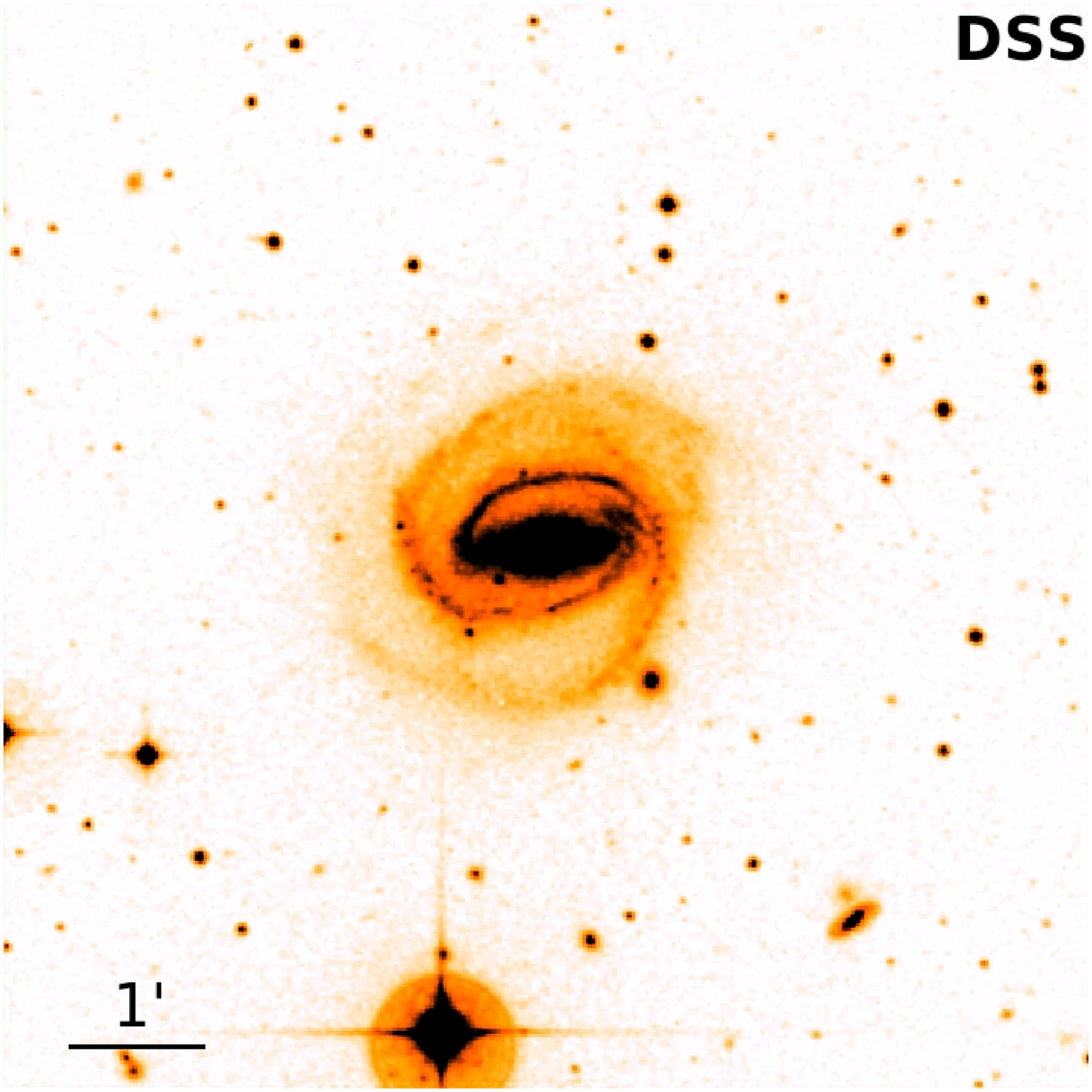}
\hspace{0.1cm} 
      \epsfxsize=5.8cm\epsfbox{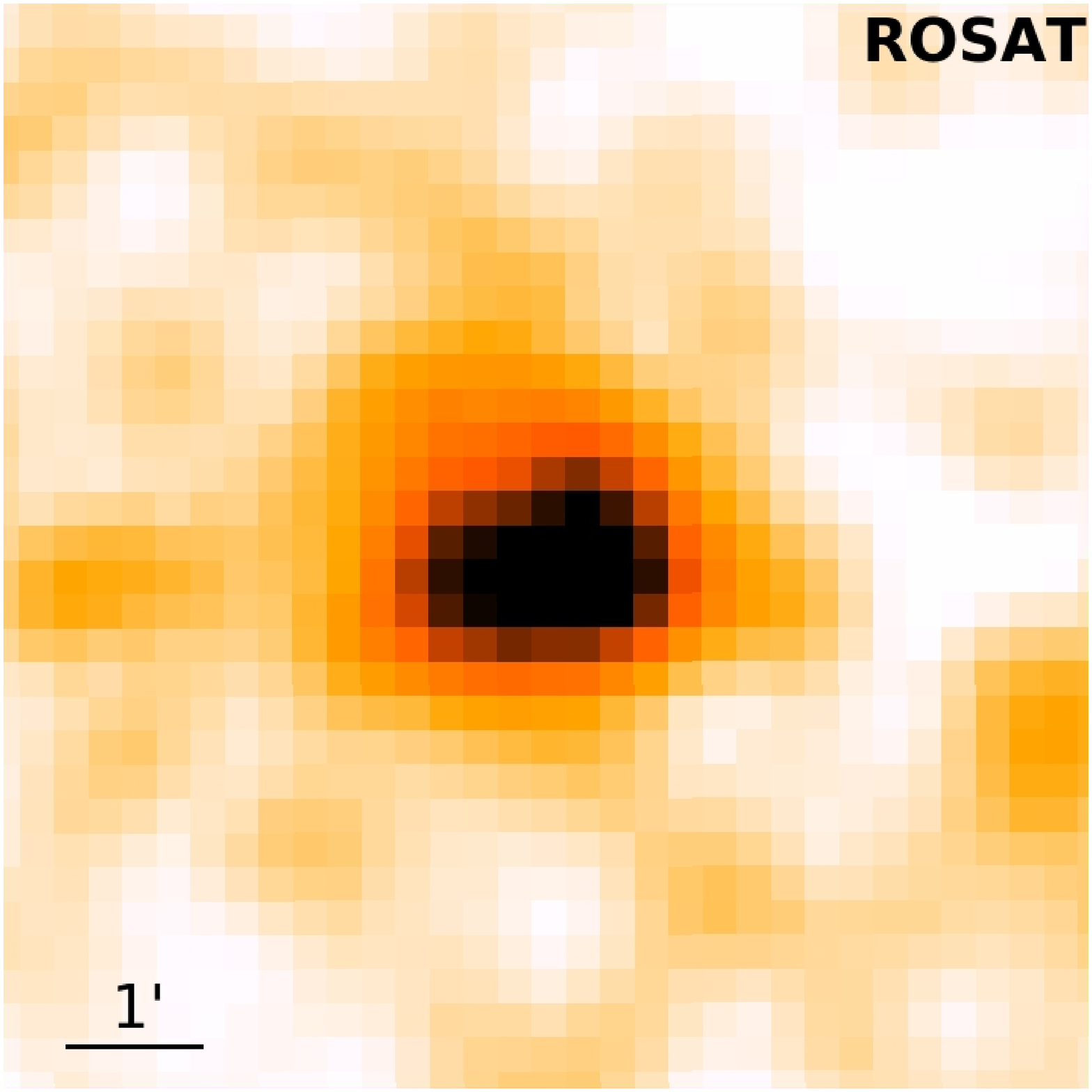}
      \caption{$8\arcmin\times8\arcmin$ ($140\times140$ kpc) images of NGC266 at near-infrared, optical, and X-ray wavelengths. The left panel shows the 2MASS K-band image of the galaxy, the middle panel depicts the DSS B-band image, and the right panel shows the $45\arcsec$ Gaussian smoothed $0.4-2.4$ keV band \textit{ROSAT} image. The X-ray emission appears to be more extended than the infrared or optical images of the galaxy, suggesting the presence of a hot X-ray corona in the dark matter halo of NGC266.}
     \label{fig:images}
  \end{center}
\end{figure*}

In this paper, we explore the extended X-ray emission associated with NGC266, using \textit{ROSAT} and \textit{Chandra} X-ray observations. NGC266 is a massive spiral galaxy (SB(rs)ab) with a very modest star-formation rate ($2.4 \  \rm{M_{\odot} \ yr^{-1} }$), hence its coronal X-ray emission is not dominated by starburst driven winds (Sect. \ref{sec:ir}). NGC266 is the dominant member of a poor group with six additional low-mass galaxies. The galaxy hosts a low-luminosity active galactic nucleus (LLAGN) \citep{doi05}.  For its distance we adopt $\rm{D}=60.3$ Mpc ($1\arcmin=17.55$ kpc).  The Galactic column density towards NGC266 is $\rm{N_H}=5.7\times10^{20} \rm{cm^{-2}}$ \citep{kalberla05}. Based on the stellar mass of the galaxy and using the results of galaxy formation simulations,  the estimated virial mass of NGC266 is $M_{\rm{200}}\sim8\times10^{12}\ \rm{M_{\odot}}$, hence its virial radius is $r_{\rm{200}}\sim410$ kpc.

\section{Data reduction}
\subsection{\textit{ROSAT}}
\textit{ROSAT} observed NGC266 in a pointed PSPC observation on January 3 1992 for 23.4 ks. The data analysis was performed with standard \textsc{ftools} packages.  Throughout the analysis, we used the  $0.4-2.4$ keV data, which offers substantial collecting area, but low instrumental background level. Indeed, the bulk of the \textit{ROSAT} background lies in the $0.1-0.4$ keV band, hence excluding this energy range significantly increases the signal-to-noise ratio for spectra that are not particularly soft. We emphasize that at energies $<$$0.4$ keV, no significant emission is expected from the hot X-ray gas due to the combined effects of the low effective area of \textit{ROSAT} and the rather high Galactic column density. 

To study the diffuse emission, we identified and excluded bright point sources. The size of the point source regions were selected to excise $\gtrsim$$95\%$ of the counts. To account for the X-ray emission associated with the LLAGN, we constructed the \textit{ROSAT} PSPC point spread function (PSF) distribution using the prescription of \citet{boese00}. The applied PSF reconstruction is the convolution of five components, whose origin and importance is discussed in \citet{hasinger92}. According to the in-flight parametrization of the PSF, the analytic PSF distributions offer a robust description of bright point sources at all off-axis radii for energies  $>$$0.188$ keV \citep{hasinger92}. 

To precisely account for all (instrumental and sky) background components, preferably local background regions should be used. Therefore, we employed a circular annulus with $4-8\arcmin$ radii around NGC266. Within this region no diffuse emission is detected from the galaxy, and the X-ray emission does not show significant radial variations, making this region suitable to calculate the background level. Finally, we mention that the vignetting effects were also corrected using the exposure map provided for the specific observation. 
  
\subsection{\textit{Chandra}}
NGC266 was observed by \textit{Chandra} on June 1 2001 (Obs ID: 1610) for 2.0 ks by ACIS-S using a $1/8$ subarray configuration with the primary goal to study the LLAGN \citep{terashima03}. Due to the small image are, these data cannot be used to explore the extended X-ray emitting components of the galaxy.

We analyzed the data with standard \textsc{ciao} software package tools (\textsc{ciao} version 4.5; 
\textsc{caldb} version 4.5.5.1). The data was reprocessed with the \textit{chandra\_repro} task. To identify bright  point sources, we ran the \textit{wavdetect} tool. The source detection sensitivity of the observation is rather high, $\sim$$10^{40}\ \rm{erg\ s^{-1}}$, hence only one point source is resolved at the center of NGC266, which is the LLAGN. 

\begin{figure*}[t]
  \begin{center}
    \leavevmode
      \epsfxsize=8.7cm\epsfbox{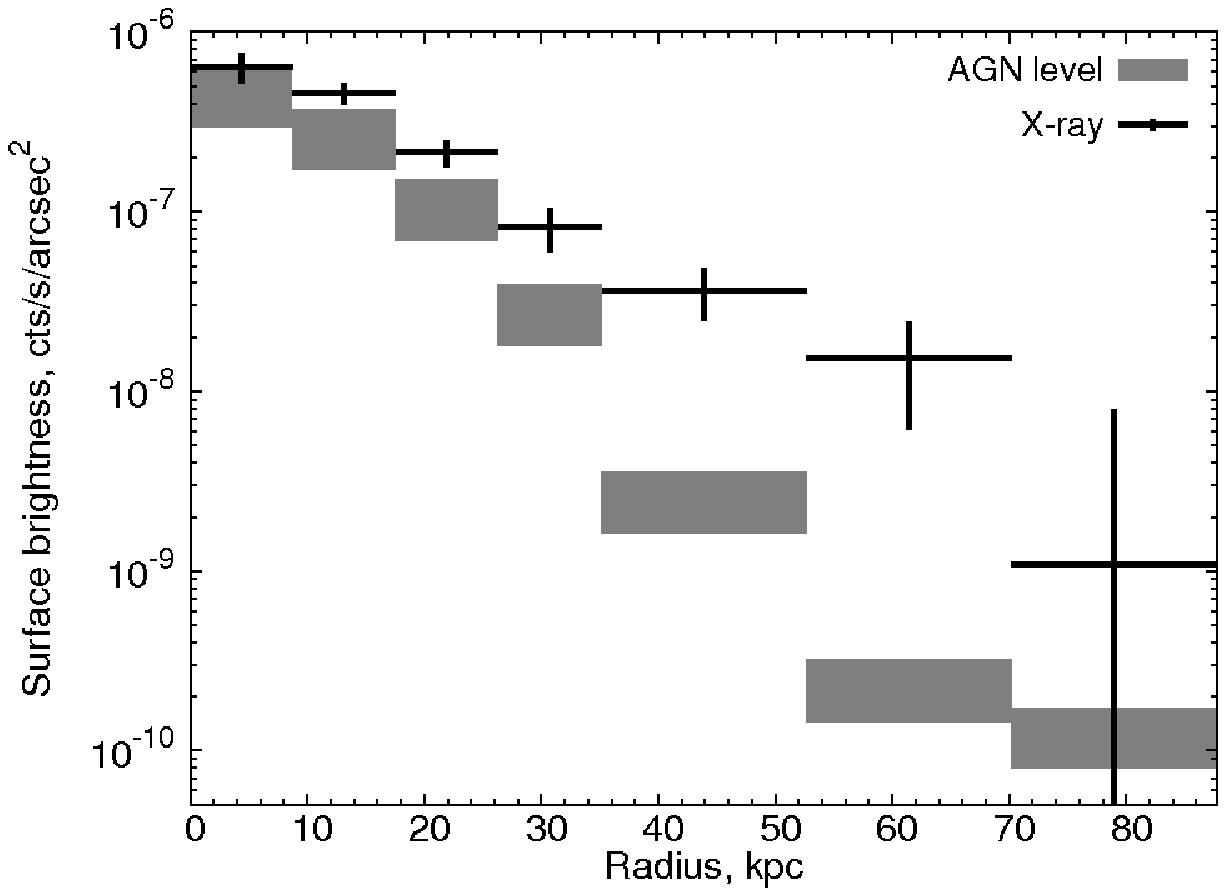}
\hspace{0.2cm} 
      \epsfxsize=8.7cm\epsfbox{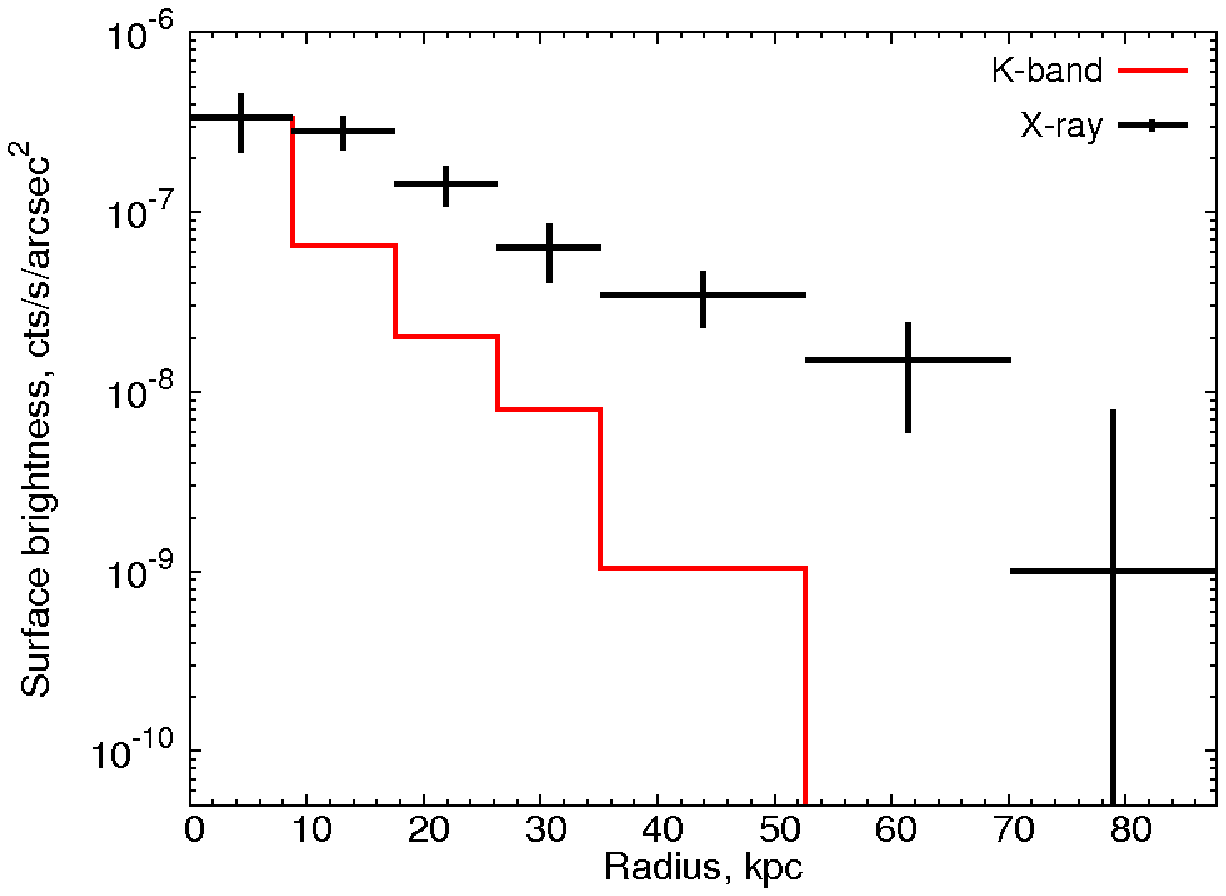}
      \caption{\textit{Left:} Surface brightness profile from \textit{ROSAT} data in the $0.4-2.4$ keV band using circular annuli. The background components are subtracted, whose level is $6.3\times10^{-8}\ \rm{cts\ s^{-1}\ arcsec^{2}}$. The shaded area shows the estimated emission associated with the LLAGN. The lower limit of the range assumes that the luminosity of the source remained constant between the \textit{ROSAT} and \textit{Chandra} observations. The upper limit is computed assuming that all counts within $0.5\arcmin$ are associated with the LLAGN. In the $2-4\arcmin$ ($\sim$$35-70$ kpc) range the  contribution of the LLAGN is in the $\sim$$3-7\%$ range. \textit{Right:}  Same as in the left panel, but the contribution of the LLAGN is subtracted assuming that its luminosity remained constant between the \textit{ROSAT} and \textit{Chandra} observations. The histogram shows the K-band light distribution, which is normalized to the innermost bin. The diffuse X-ray emission extends to at least $4\arcmin$ ($\sim$$70$ kpc), while the bulk of the K-band light is confined to within  $\sim$$1.5\arcmin$ ($\sim$$26 $ kpc).}
     \label{fig:profile}
  \end{center}
\end{figure*}

\subsection{Near-infrared data}
We used the K-band images of the Two Micron All Sky Survey (2MASS) to trace the stellar light \citep{jarrett03}. Since the 2MASS K-band images of NGC266 are not background subtracted, we used nearby regions away from the galaxy to estimate the background level. Bright foreground stars were visually identified and removed.

\section{Results}
\subsection{Overall properties of NGC266}
\label{sec:ir}
We derive the total stellar mass of NGC266 from its K-band luminosity. To derive the K-band luminosity, we rely on the  apparent 2MASS K-band magnitude of the galaxy ($m_{\rm{K}}=8.673$ mag) and use the absolute K-band magnitude of the Sun ($M_{\rm{K,\odot}}=3.28$ mag). The K-band luminosity of NGC266 is  $L_K=2.5\times10^{11} \ \rm{L_{K,\odot}}$. We convert this value to stellar mass using the K-band mass-to-light ratio of $M_{\star}/L_{K}=0.80$, which we compute from the results of galaxy evolution modeling \citep{bell03} and the $B-V=0.82$ color index. Hence, the stellar mass of NGC266 is $2.0\times10^{11}\ \rm{M_{\odot}}$. 

The star-formation rate (SFR) of NGC266 is computed from the total infrared luminosity employing the \citet{kennicutt98} relation. To derive the far-infrared fluxes, we follow the methodology of \citet{meurer99} and rely on the $60\ \mu$m and $100\ \mu$m fluxes obtained by the \textit{Infrared Astronomical Satellite}. From the far-infrared luminosity ($L_{\rm{FIR}}=3.0\times 10^{43}\ \rm{erg \ s^{-1}}$), we derive the total infrared luminosity ($L_{\rm{TIR}}=5.3 \times 10^{43}\ \rm{erg \ s^{-1}}$) using the conversion of $L_{\rm{TIR}}=1.75 L_{\rm{FIR}}$ \citep{calzetti00}. Hence, the SFR of NGC266 is  $2.4\  \rm{M_{\odot}\ yr^{-1} }$, which is a rather low value given the stellar mass of the galaxy. 

\subsection{Multi-wavelength images}
In Figure \ref{fig:images} we depict an $8\arcmin\times8\arcmin$ ($140\times140$ kpc) region of NGC266 in near-infrared, optical, and X-ray wavelengths. The 2MASS K-band image traces the old stellar population, hence this image highlights the bulge of the galaxy. In the B-band optical image, taken from the Digitized Sky Survey (DSS), the spiral arms are prominent. The image also shows that NGC266 is seen approximately face-on. Both the 2MASS and DSS images indicate the relaxed and undisturbed nature of the galaxy. The right panel shows the $0.4-2.4$ keV \textit{ROSAT} image smoothed with a $45\arcsec$  Gaussian kernel. The X-ray image exhibits a bright central region and shows the presence of large-scale extended emission. This emission is more extended than  the stellar light, which hints at the presence of a hot X-ray corona surrounding the galaxy.  However, NGC266 hosts an LLAGN at its nucleus, whose emission could contribute to the observed diffuse emission due to the broad point spread function (PSF) of \textit{ROSAT}. To assess the importance of the LLAGN in the extended emission, we quantitively investigate the X-ray and K-band surface brightness distribution of NGC266. 

\begin{table*}
\caption{The parameters of the extended hot gas coronae in the $(0.05-0.15)r_{200}$ radial range.}
\begin{minipage}{18cm}
\renewcommand{\arraystretch}{1.4}
\centering
\begin{tabular}{c c c c c c c c}
\hline 
Galaxy&  $L_{\rm{0.5-2keV,abs}}$ & $L_{\rm{bol}}$  &$M_{\rm{gas}}$ & $n_{\rm{e}}$  & $t_{\rm{cool}}$ \\ 
 &  $\rm{erg\ s^{-1}}$ & $\rm{erg \ s^{-1}}$ &$M_{\rm{\odot}}$ & $\rm{cm^{-3}}$ & Gyr  \\ 
\hline
NGC266 &$(1.7\pm1.3)\times10^{40}$ & $(4.3\pm0.8)\times10^{40}$ &  $(9.1\pm0.9)\times10^{9}$ & $(4.0\pm0.4)\times10^{-4}$ & $38\pm4$ \\
\hline
NGC1961$^{\dagger}$ & $(2.0\pm0.6)\times10^{40}$ &  $(5.8\pm1.7)\times10^{40}$  & $(1.2\pm0.2)\times10^{10}$ & $(3.5\pm0.5)\times10^{-4}$ & $45\pm6$ \\
NGC6753$^{\dagger}$ & $(2.5\pm0.4)\times10^{40}$ &   $(6.3\pm0.9)\times10^{40}$ & $(1.1\pm0.1)\times10^{10}$ &$(3.9\pm0.3)\times10^{-4}$ & $51\pm4$ \\
\hline \\
\end{tabular} 
\end{minipage}
$^{\dagger}$ Values taken from \citet{bogdan13}.
\label{tab:values}
\end{table*}  

\subsection{Surface brightness distribution}
To study the X-ray emission associated with NGC266, we construct a surface brightness profile from the $0.4-2.4$ keV band \textit{ROSAT} image. The profile is extracted from circular annuli centered on the galaxy nucleus. The left panel of Figure \ref{fig:profile} shows the surface brightness profile, which reveals that the diffuse emission extends to at least $4\arcmin$ ($\sim70$ kpc). Note that the associated with the LLAGN is not excluded. To estimate the contribution of the LLAGN, we extract the \textit{ROSAT} PSF at $1.1$ keV  and at $23\arcmin$ off-axis, the approximate location of NGC266. The particular energy choice does not influence our conclusions, since the \textit{ROSAT} PSF is only weakly dependent on the energy \citep{boese00}.  According to the PSF distribution, $\sim$$95\%$ of the  source counts from the LLAGN should be within $2\arcmin$ radius. To estimate the luminosity of the LLAGN, we use two approaches. First, we use the \textit{Chandra} data to derive the expected number of counts detected by \textit{ROSAT}. With the $2$ ks \textit{Chandra} exposure we detect $40.5$ net counts in the $0.5-8$ keV band, which corresponds to $62.4$ \textit{ROSAT} counts in the $0.4-2.4$ keV band, if we adopt a power law spectrum with slope of  $\Gamma=1.4$ and Galactic column density. This method assumes that the luminosity of the LLAGN did not vary between the two observations. However, the X-ray source  could have been more luminous at the epoch of the \textit{ROSAT} observation. Hence as a second approach, we assume that all detected \textit{ROSAT} counts within the central $30\arcsec$ are associated with the LLAGN. Under this extreme condition, the point source was $\sim$$2.1$ times brighter, implying that $132.7$ \textit{ROSAT} counts are associated with it.  Convolving the number of estimated source counts with the PSF distribution, we show the estimated contribution of the LLAGN to the diffuse emission with shaded area in Figure \ref{fig:profile}. The profile demonstrates that beyond $\sim$$1.5\arcmin$ ($\sim$$26$ kpc), the contribution of the LLAGN does not play a major role. Specifically, in the $2-4\arcmin$ ($\sim$$35-70$ kpc) region no more than $\sim$$3-7\%$ of the detected counts are associated with the bright nuclear source. Thus, the large-scale extended emission around NGC266 cannot be explained by the LLAGN. Moreover, the left panel of Figure \ref{fig:profile} demonstrates that the extended emission and the PSF exhibit markedly different distributions, implying that the extended emission cannot be merely attributed to the population of unresolved point sources associated with NGC266. 

In the right panel of Figure \ref{fig:profile}, we show  the actual surface brightness profile of the diffuse X-ray emission. From this profile  we subtracted the counts associated with the LLAGN, assuming that its luminosity did not change between the \textit{ROSAT} and \textit{Chandra} observations. To trace the stellar light, we extract the radial profile of the  2MASS K-band image using the same annuli as for the X-ray profile. The K-band profile is normalized to match the level of X-ray emission in the innermost bin. As suggested by the images (Figure \ref{fig:images}), the surface brightness profile demonstrates that the X-ray light associated with NGC266 extends well beyond the stellar light. While $\sim$$90\%$ of the K-band emission is confined to within $1.5\arcmin$ ($\sim$$26$ kpc), the diffuse X-ray emission extends to at least $4\arcmin$ ($\sim$$70$ kpc), implying that it is not  associated with the X-ray emitting components of the stellar body. The radial distribution of the diffuse emission around NGC266 is similar to those obtained in the two other  well-studied X-ray coronae in massive spirals \citep{anderson11,bogdan13}. To further study the spatial distribution of the diffuse emission, we divided NGC266 into three  azimuthal sectors with opening angles of $120\degr$. The azimuthal profiles in the different sectors agree with each other within statistical uncertainties, indicating that the extended emission has a fairly uniform distribution. However, due to the low number of source counts in the outer regions, followup observations are required to map the spatial distribution of the emission in detail.

\subsection{X-ray hardness ratios}
To unveil the nature of the diffuse emission beyond the optical extent of NGC266, it would be desirable to study its X-ray spectrum. However, in the $2-4\arcmin$ region, \textit{ROSAT} only detects  $\sim$$40$ net counts, which are not sufficient for spectral analysis. Therefore, we derive X-ray hardness ratios that could differentiate between the two potential origins of the emission: (i) hot gaseous X-ray corona in the dark matter halo of NGC266, or (ii) a multitude of unresolved point sources, presumably originating from (background) AGN. Defining the hardness ratio ($\rm{HR}$) as the ratio of net counts $ (1.2-2.4\ \rm{keV})/(0.4-1.2\ \rm{keV})$, we expect very different HRs in the two scenarios. Assuming a thermal spectrum with $kT=0.6$ keV, a metallicity of $0.2$ Solar, and Galactic column density, we expect $\rm{HR}=0.13$. Note that such a spectrum was observed for the X-ray coronae of NGC1961 and NGC6753 \citep{bogdan13}. If (background) AGN dominate, whose spectrum can be described with a power law model with slope of $\Gamma=1.4$ and Galactic column density, we expect $\rm{HR}=0.64$. 

Within the central $2\arcmin$ region, we obtain $\rm{HR}=0.78\pm0.14$, which  is in good agreement with a hard power law spectrum.  Indeed, the central region is dominated by the LLAGN, while the population of unresolved LMXB and HMXBs also plays a role. Their spectra can be described with similar power law models. In the $2-4\arcmin$ region, where the contribution of stellar light and the LLAGN is negligible, we obtain $\rm{HR}=0.20\pm0.18$. This indicates that the emission most likely originates from a  soft X-ray emitting component, and this value is consistent with a hot gaseous emission with a temperature less than $1$ keV. However, given the large statistical uncertainties, the contribution of a harder component, for example a population of unresolved cosmic X-ray background sources, cannot be excluded. Nevertheless, the X-ray hardness ratios indicate that the bulk of the diffuse emission beyond the stellar body of NGC266 is consistent with that from a gaseous X-ray corona. 

\subsection{Estimated gas parameters}
Measuring the properties of the hot X-ray gas in the dark matter halos of massive spirals can place major constraints on galaxy formation models, hence on the physical processes that influence galaxy evolution. To facilitate these efforts, we also derive the  gas parameters for NGC266. Since accurate X-ray determinations of the temperature and metallicity are not feasible,  we adopt gas parameters, which were measured for the hot X-ray coronae around NGC1961 and NGC6753 \citep{bogdan13}. Namely, we assume a gas temperature of $kT=0.6$ keV and a metallicity of $0.2$ Solar.

We compute the gas parameters in the $(0.05-0.15)r_{200}$ region, which corresponds to an annulus from $1.17-3.5\arcmin$ ($\sim$$20.5-61.5$ kpc). Within this region we detect $79.0\pm13.9$ net counts after accounting for the emission associated with the LLAGN. In this computation we assumed that the luminosity of the LLAGN did not change between the \textit{ROSAT} and \textit{Chandra} observations. Assuming the above-described thermal spectrum, the detected counts correspond to an absorbed $0.5-2$ keV luminosity of $L_{\rm{0.5-2keV,abs}}=(1.7\pm0.3)\times10^{40}\ \rm{erg \ s^{-1}}$. The bolometric luminosity of the gas in the $(0.05-0.15)r_{200}$ region is $L_{\rm{bol}}=(4.3\pm0.8)\times10^{40}\ \rm{erg \ s^{-1}}$. Using the volume from which the source flux is extracted and assuming that the emission can be described with a thermal plasma emission model (\textsc{apec} in \textsc{Xspec}), we estimate that the mean electron density of the gas is $n_{\rm{e}}=(4.0\pm0.4) \times 10^{-4} \ \rm{cm^{-3}}$. Assuming a volume filling factor of unity and a constant gas density, the total  gas mass is $M_{\rm{gas}}=(9.1\pm0.9)\times10^9\ \rm{M_{\odot}}$. The cooling time of the gas is estimated from  $t_{\rm{cool}}=(3kT)/(n_e\Lambda(T))$, where $\Lambda(T)$ is the cooling function, and we obtain $t_{\rm{cool}}=38\pm4$ Gyrs. These values are listed in Table \ref{tab:values}, where we also include the gas parameters derived for NGC1961 and NGC6753. The comparison reveals that the gas parameters are similar in all three galaxies. 

\section{Discussion}
Following \citet{bogdan13} we investigate if starburst driven winds could be responsible for the observed extended emission around NGC266. According to theoretical considerations, extraplanar X-ray emission driven by starburst winds is expected if the area specific supernova (SN) rate is $\gtrsim$$25\ \rm{SN\ Myr^{-1} kpc^{-2}}$ \citep{strickland04}. We derive the SN rate from the total far-infrared luminosity $R_{\rm{SN}}=0.2 L_{\rm{FIR}}\ 10^{11} L_{\rm{\odot}}$ \citep{heckman90}, and obtain $R_{\rm{SN}}=0.015\ \rm{yr^{-1}}$. Based on this value and the size of the $D_{\rm{25}}$ major axis diameter, the area specific SN rate of NGC266 is $\sim$$5.7\ \rm{SN\ Myr^{-1} kpc^{-2}}$, which falls short of the critical value. Thus, NGC266 is unlikely to drive starburst winds at the present epoch. 

The detection of a luminous X-ray corona around NGC266 raises the question if other coronae could be detected  based on archival \textit{ROSAT} data. In principle, further detections are feasible with \textit{ROSAT}, provided that sufficiently deep pointed observations are available. However, in the local Universe ($\lesssim$$100$ Mpc) most massive normal spiral galaxies were only observed with the \textit{ROSAT} All Sky Survey or with short pointed PSPC observations, which data are not suitable for detailed analysis. Therefore, to extend the sample of massive spirals with luminous coronae, observations with present-day X-ray telescopes are indispensable. 

While the present  data demonstrate the presence of an extended X-ray emitting component around NGC266, a moderately deep \textit{Chandra} observation could achieve the following goals: (i) conclusively determine the nature of the extended emission; (ii) assess the X-ray luminosity of the LLAGN; (iii)  measure the gas temperature and metallicity, and hence derive accurate gas parameters; (iv) build a precise radial profile, and hence estimate the total confined gas mass, and compute the baryon budget of the galaxy. 

NGC266 is the fourth spiral galaxy with a luminous X-ray corona detected beyond its optical radius. Although the increasing number of detections is encouraging, the present sample does not yet allow us to draw general conclusions about the properties of X-ray coronae.  The detected galaxies have a narrow range of stellar masses. Therefore, it is essential to characterize the presently detected  coronae in more detail and to explore further -- preferably lower mass -- galaxies. At present, galaxy formation simulations are fairly mature, incorporating supernova and AGN feedback, setting the stage to confront the observations with realistic simulations. Such comparisons offer a unique possibility to constrain the physical processes, in particular the feedback effects, that play a crucial role in the evolution of galaxies from high redshifts to the present-day Universe. 

\bigskip
\begin{small}
\noindent
We thank the anonymous referee for helpful comments. \'AB thanks Alexey Vikhlinin for helpful discussions about the \textit{ROSAT} PSF. This research has made use of \textit{Chandra}  data provided by the CXC. This publication makes use of data products from the 2MASS, which is a joint project of the University of Massachusetts and the Infrared Processing and Analysis Center/California Institute of Technology, funded by NASA and NSF. The authors acknowledge the usage of the HyperLeda database. \'AB acknowledges support provided by NASA through Einstein Postdoctoral Fellowship grant number PF1-120081 awarded by the CXC, which is operated by the Smithsonian Astrophysical Observatory for NASA under contract NAS8-03060. WF and CJ acknowledge support from the Smithsonian Institution. 
\end{small}

\end{document}